\documentclass{appolb}
\usepackage{graphicx}

\begin{document}
\title{ Unresolved issues in the search for eta-mesic nuclei%
}
\author{N. G. Kelkar
\address{Dept. de Fisica, Univ. de los Andes, Cra 1E, 18A-10, Bogota, Colombia}
}
\maketitle
\begin{abstract}
Even if the theoretical definition of an unstable state is straightforward, its 
experimental identification often depends on the method used in the analysis and 
extraction of data. A good example is the case of eta mesic nuclei where strong 
hints of their existence led to about three decades of extensive theoretical and 
experimental searches. Considering the still undecided status of these states and 
the limitations in the understanding of the eta-nucleon as well as the eta-nucleus 
interaction, the present article tries to look back at some unresolved problems 
in the production mechanism and final state interaction of the eta mesons and nuclei. 
An unconventional perspective which provides a physical insight into the nature of 
the eta-nucleus interaction is also presented using quantum time concepts.
\end{abstract}
\PACS{21.85.+d,13.60.Le,13.75.-n,03.65.Xp}
  
\section{Introduction}
The strong nuclear interaction which is usually understood as the interaction 
between two nucleons is by now quite well known. However, the interaction is 
mediated by mesons which are also strongly interacting objects and it makes sense 
to have a good understanding of the meson-nucleon and 
especially the meson-nucleus interaction too. Hadron-nuclear interactions in general 
can be probed either via scattering experiments or the study of bound 
states. It has been possible to investigate the pion-nucleus interaction 
through elastic scattering experiments, but the same is not true of the heavier 
eta meson due to its being extremely short lived. 
Due to the non-availability of eta beams, the eta-nucleon ($\eta$-N) 
interaction can only
be deduced via the eta-nucleon or eta-nucleus interaction in the final state.
Fortunately, its interaction with the nucleon in the $s$-wave 
(which proceeds through the formation of an N$^*$(1535) resonance) is 
attractive and leads to the possibility of forming and studying 
unstable bound states of eta mesons and nuclei \cite{haiderliu} 
(see also \cite{wilkrev,machrev,ourrev} for an extensive list of references). 
After about 25 years of investigations in this field, 
a lot of progress has been made and
some evidence for the existence of such states exists, however, 
there still does not exist a
general agreement and a final word on the strength of the eta-nucleon and 
eta-nucleus interaction. 

The eta meson being heavy with a mass around 547 MeV, 
reactions producing eta mesons on nuclei involve large momentum 
transfers to the nuclei involved. This fact gives rise to the possibility of 
reaction mechanisms \cite{lagetlecol} which go beyond the one step process 
where a single nucleon gets 
excited to the N$^*$ resonance which eventually decays into an $\eta$ and a nucleon 
\cite{santrabk}. The $\eta$-N and hence the $\eta$-nucleus interaction can 
be deduced only from reactions 
where the $\eta$ is produced in the final state. Such a deduction then depends 
on the theoretical model used for the reaction mechanism for production as 
well as the final state interaction of the $\eta$ meson with the nucleus.
A theoretical prediction for the existence of an eta-mesic unstable nuclear state 
which requires the strength of the $\eta$-N interaction as an input, 
thus depends indirectly on the 
correctness of the reaction mechanisms describing the production reactions. 
Besides this, the strong effects of the $\eta$-nucleus interaction are 
prominent near the threshold of the eta producing reactions where the off-shell 
rescattering of the eta and the nucleus becomes important. Methods based on the 
extraction of the eta-nucleus scattering length (which is further used to 
comment on the possible existence of eta mesic nuclei) involving on-shell 
approximations can indeed lead to quite different conclusions as compared to 
few body calculations of the same. Here, an attempt to compare such  
discrepancies and look back at the deficiencies in explaining the meson 
production data will be made. The next section begins with a brief introduction 
to the methods of identifying unstable states focussing especially on the 
application of a quantum time concept 
introduced by Wigner and Eisenbud \cite{wigner} 
and modified recently for locating eta mesic nuclear states \cite{meprl}. 
The sections which follow, discuss the experimental searches, the limitations 
of some of the theoretical approaches and possibilities for the future.

\section{Identification of unstable mesic states}
Baryon resonances are usually identified by 
performing a partial wave analysis of the elastic meson
baryon scattering data and obtaining 
the energy dependent amplitude (or transition matrix) by fitting cross section data. 
Resonances are then determined by locating the poles of the S-matrix
on the unphysical sheet and studying the Argand diagrams of this complex transition 
matrix. With the eta meson being extremely short lived ($\tau$ $\sim$ 
10$^{-18}$ s), 
the possibility of performing such analyses with 
elastic $\eta$-N or eta-nucleus elastic scattering data 
does not exist. Hence the next best thing to do in a theoretical search 
is to use the information on the 
$\eta$-N interaction obtained from models fitting the 
$\eta$ meson producing reaction data and use it as an input to theoretically 
construct a complex eta-nucleus elastic scattering matrix. 
The poles of this matrix in the complex energy/momentum plane can then be used 
to infer on the existence of unstable states. 
The complex energy $E$ is related to the
complex momentum $p$ as $E = p^2 /2\mu$ where $\mu$ is the $\eta$-nucleus
reduced mass. The physical and unphysical sheets correspond to
$\Im$m $p >$ 0 and $<$ 0 respectively.
Denoting $\Re$e $p = p_{\!_R}$ and $\Im$m $p = p_{\!_I}$,
$$E = {1 \over 2\mu} \, (\,p_{\!_R}^2 \, -\, p_{\!_I}^2 \, + 2\, i\, p_{\!_R} 
\,p_{\!_I})$$
and hence, $\Re e\, E\, = (p_{\!_R}^2 - p_{\!_I}^2 )/2\mu$ and
$\Im m \, E\, = p_{\!_R} p_{\!_I}/ \mu$. 
For the existence of a bound (or quasibound)
state, the requirement is $\Re e \,E < 0$.
This means that for a quasibound state to exist, $p_{\!_R}^2 < p_{\!_I}^2$ and
for $p_{\!_I} > 0$, the pole, of the type $-|{\cal E}| - i \Gamma/2$ 
should lie in the second quadrant
(see Figure 5 of Ref. \cite{ourrev}) of the complex $p$ plane
above the diagonal which divides this quadrant into two.
As $p_{\!_R} \to 0$, the pole lies on the positive imaginary $p$ axis 
and corresponds to a bound state. The virtual state pole lies on the negative imaginary 
$p$ axis. 
Resonances are defined as the states on the unphysical sheet 
($\Im$m $p < 0$) with $\Re e \,E > 0$, i.e. a pole of the type 
$|{\cal E}| - i \Gamma/2$ . Quasivirtual states lie on the unphysical sheet too, but 
with a pole like  $-|{\cal E}| + i \Gamma/2$, they lead to an exponential growth 
and not decay and hence are not physical unstable states. 

\subsection{Wigner's time delay and the dwell time method} 
The $\eta$-nucleus transition matrix can also be used to evaluate the so-called 
time delay (a concept initially introduced by Wigner and Eisenbud \cite{wigner} 
and elaborated 
by Smith and many others later \cite{smith,us}) which is large and positive with a
typical Lorentzian (Breit-Wigner type) shape at positive and negative energies 
for resonances and quasibound states respectively. Virtual and quasivirtual states 
lead to negative delay times \cite{ourjphys}. A variation of this concept, namely, 
the dwell time delay which is useful for 
identifying unstable states near threshold was introduced in \cite{meprl}. 
The relation between the dwell and
phase time delay, $\tau_D(E)$ and $\tau_{\phi}(E)$ respectively, 
in scattering was found to be
\begin{equation}\label{fin}
\tau_D(E) \,=\, \tau_{\phi}(E)\, +\, \hbar \,
\mu\, [t_R / \pi] \,\,dk / dE\,.
\end{equation}
where the phase time delay is given in terms of
the $S$-matrix as $\tau_{\phi}(E) \,=\, 
\Re e [ -i \hbar (\,S^{-1} {dS / dE}\,)\,]$, with, 
$S = 1 \,-\, i \mu\,k \,(t_R \,+\, it_I)/\pi$. 

Since time delay is the difference in the time spent 
by the scattering objects in a given region with and without interaction, 
an attractive interaction would be expected to ``delay" the process whereas a 
repulsive one would cause the time spent with interaction to be smaller and 
hence cause the difference to be negative.
Thus a large positive time delay 
signals the existence of an unstable bound (quasibound) state or a 
resonance which is formed, propagates and decays delaying the process.

\subsection{$\eta$-$^3$He and $\eta$-$^4$He mesic states}
A time delay analysis in \cite{ourjphys} led the authors 
to conclude that only small $\eta$-N scattering lengths favour the formation of 
light quasibound eta mesic nuclei. 
A large scattering length such as $a_{\eta N} = (0.88, 0.41)$ fm which in principle 
is the value which reproduces the p d $\to$ $^3$He $\eta$ 
reaction data well \cite{wekanchannpa} 
leads to a (positive energy) $\eta$-$^3$He resonance and a 
strong repulsion near the threshold for $\eta$-$^4$He 
states (in agreement with absence of quasibound states as found in \cite{moskal4He}). 
A quasibound state such as
that claimed in \cite{pfeiffer} and later withdrawn \cite{pheron} 
can be formed only with
a small $\eta$-N scattering length. In fact, the authors \cite{meprl,ourjphys} 
noticed a movement of the poles from a relatively deep quasibound state 
at (- 5 - $i$ 8) MeV to one 
near threshold at (0 - $i$ 1.95) MeV 
and then a resonance at (0.5 + $i$ 0.65) MeV, 
for increasing values of the scattering length, 
$a_{\eta N}$ = (0.28 + $i$0.19) fm, (0.51 + $i$ 0.26) fm 
and (0.88 + $i$ 0.41) fm respectively. 
 
One of the hints for the possible existence of eta mesic $^3$He was 
the rapid increase of the magnitude of the $s$-wave amplitude 
in the p d $\to$ $^3$He $\eta$ reaction near threshold. 
In \cite{wilkinplb} the authors showed that the phase of the $s$-wave amplitude 
varies strongly near threshold too. 
Considering contributions of the $s$ and $p$ waves, the authors found that the 
angular distribution of this reaction is sensitive to the $s$-$p$ interference. 
They associated the sharp rise 
in the total cross sections with the existence of  
a pole corresponding to either a quasibound or a quasivirtual state very close 
to threshold ($Q_0 = (-0.30 \pm 0.15_{\small stat} \pm 0.04_{\small syst}) \pm 
i (0.21 \pm 0.29_{\small stat} \pm 0.06_{\small syst})$ MeV).

In a much earlier work \cite{wycech}, within a Watson-like multiple scattering 
formalism, the authors tried to simultaneously analyse the p d $\to$ $^3$He $\eta$ 
as well as the d d $\to$ $^4$He $\eta$ data in order to investigate the existence 
of eta-mesic helium nuclei. With the $\eta$-N interaction not being well-known, 
they studied the existence of such states for various values of the 
$\eta$-N scattering length. 
The authors concluded that though $\eta$-$^3$He 
quasibound states would be less likely to exist, $\eta$-$^4$He states could possibly 
exist. 
In passing, we also mention Ref. \cite{willis}, where, 
based on a scattering length approximation for the final state 
interaction and with a simultaneous fit to the d d $\to$ $^4$He $\eta$ and 
p d $\to$ $^3$He $\eta$ data the authors 
determined $a_{ \eta ^3He} \simeq (-2.3 + i3.2)$ fm ,
and $a_{\eta ^4He} \simeq (-2.2+ i1.1)$ fm indicating the existence of both 
$\eta$-$^3$He and $\eta$-$^4$He states.  
Conclusions about the existence of eta mesic states based on scattering 
length approaches for the final state should however be taken with some caution 
as will be discussed in the next section. 
Some hope of shedding light on the existence of these light nuclear 
states lies in the ongoing 
efforts made by the WASA collaboration at COSY \cite{wasa1,wasa2,wasa3,wasa4,
moskalfew}. 

\section{Analyses of $\eta$ producing reactions}
The basic ingredients of any theoretical analysis of an $\eta$ meson producing 
reaction on nuclei are (i) a model for the reaction mechanism to produce 
an $\eta$ in the final state (ii) a framework to incorporate the $\eta$-nucleus 
final state interaction (FSI) and (iii) information on the 
interaction of the $\eta$ meson with the nucleons within the nucleus. 
The large momentum transfer to the nucleus involved in these reactions rules out 
the possibility of using a simple model where the $\eta$ is produced 
directly from the decay of the $N^*$ resonance in a one step process. Since data 
on $\eta$ production have confirmed the existence of strong FSI effects near 
threshold, it is also important to notice that the $\eta$ meson can in 
principle be produced off-shell and brought on-shell (after several 
rescatterings) by its FSI with the nucleus. Below, we try to analyse the 
deficiencies in literature in the treatment of these three main ingredients 
in the search of eta mesic nuclei. 
\subsection{Two step models of $\eta$ -helium production}
With the eta mass being around 547 MeV, the momentum transferred to the 
nucleus is large. Laget and LeColley noticed the need for a three-body 
mechanism where the momentum is shared by the nucleons in the nucleus. Indeed, 
they found that \cite{lagetlecol} a one-body mechanism underestimated the 
cross sections by 2-3 orders of magnitude. Faeldt and Wilkin \cite{falwilk} 
found good agreement with the threshold data on the p d $\to$ $^3$He $\eta$ 
reaction using the so called two step model where one nucleon in the deuteron 
first interacts with the incident proton to produce an intermediate pion which 
eventually produces the eta meson via the $\pi$ N $\to$ $\eta$ N reaction. 
The transition amplitude for the two step process can be then written as
\cite{wekanchannpa}, 
\begin{eqnarray}\label{born}
< |T_{pd \rightarrow ^3{\rm He}\,\eta}| >=i \int {d\vec{p_1}\over (2\pi)^3}
{d\vec{p_2}\over (2\pi)^3} \sum_{int\,m's} <p n \,|\,d>
\,<\pi\,d |T_{pp\, \rightarrow\, \pi\, d}| p\,p>
\\ \nonumber
 \times{1\over (k_\pi^2-m_\pi^2+i\epsilon)}
\, <\eta\,p \,|\,T_{\pi N \rightarrow \eta p}\,| \pi\,N>
\,\,<\,^3{\rm He}\,|\,p\,d>\,,
\end{eqnarray} 
where the sum runs over the spin projections of the intermediate
off-shell particles and $k_{\pi}$ is the four momentum of the intermediate pion.
Motivated by the success of this model, in \cite{wekanchannpa} the final state 
$\eta$-$^3$He interaction was incorporated using few body equations which took 
into account the off shell rescattering of the $\eta$ from the nucleus. Thus 
\begin{eqnarray}\label{tmat2}
&&< \Psi_{f} (\vec{k_{\eta}})|T_{pd \rightarrow ^3{\rm He}\,\eta}| 
\Psi_i(\vec {k_p}) > \,
 =\, <\,\vec{k_\eta} ;  m_3|\, T_{p d \rightarrow \,^3{\rm He} \eta}|
\,\vec{k_p} ;  m_1 \, m_2\,> + \\ \nonumber
&&\sum_{m_3^\prime} \int { d\vec {q} \over (2\pi)^3} {<\vec{k_\eta}
; m_3|         
T_{\eta\, ^3{\rm He}} | \vec{q} ; m_3^\prime> \over E(k_\eta)\,
- \,E(q)\,
+\, i\epsilon}\,
<\vec{q} ; m_3^\prime\,| T_{p d \rightarrow \,^3{\rm He} \,\eta}
|\vec{k_p}; m_1\, m_2>.
\end{eqnarray}
The intermediate $\pi$ N $\to$ $\eta$ N process was described by a coupled 
channel off shell t-matrix. Though such a model could reproduce the threshold 
total cross sections as well as the isotropic angular distributions, it was 
later found \cite{ourstenandcom} to 
produce backward peaked angular distributions at high energies in complete 
disagreement with data. Indeed, backward peaked angular distributions 
seem to be a common feature of high momentum transfer reactions 
(like for example, p d $\to$ $^3$He $\omega$, p d $\to$ 
$^3$H$_{\Lambda}$ $K^+$) described 
within a two step model \cite{otherooltas}. Improved calculations including 
coupled channel effects arising for example from the p d $\to$ $^3$He $\pi^0$ 
reaction, the interaction of the off-shell intermediate 
particles with the inclusion of higher partial waves could lead to a better
understanding of the problem. The latter is supported by \cite{wilkinplb} 
where the authors found the angular distributions sensitive to the $s$-$p$ 
wave interference.      

\subsection{On-shell approximations}
In a proper description of the FSI between the $\eta$ and the nucleus, one
must consider the fact that the $\eta$ meson can also be produced
off the mass shell and eventually brought on-shell due to its
interaction with the nucleus. The half off-shell transition matrix 
can be written using few body equations. In \cite{wekanchannpa,weneelamBe, 
garcilazo} the off-shell FSI included using few body equations was found 
to be important for the p d $\to$ $^3$He $\eta$, p $^6$Li $\to$ $^7$Be $\eta$ 
and p n $\to$ d $\eta$ reactions. The eta nucleus scattering lengths deduced 
from such off shell t-matrices are also found to be quite different 
\cite{ourrev} from those using a rather simple on shell approximation 
with the FSI amplitude given by
\begin{equation}
f \simeq {f_B \over 1 - i k a_{\eta N}} 
\end{equation}
with $f_B$ fitted to reproduce the right magnitude of the data. 
It is surely tempting to use a simple approximation as the above expression 
to extract eta-nucleus scattering lengths 
from eta production data and comment on the existence 
of eta mesic states based on the magnitude and sign of the scattering 
length. However, such conclusions could indeed be way away from reality. 

\subsection{The $\eta$-nucleon interaction}
We started off with the aim of understanding the eta meson - nucleon 
interaction and having obtained indications of its attractive nature 
we set out on a search of its exotic nuclear states, namely, the 
eta mesic nuclei. However, extensive investigations of hadron and photon 
induced $\eta$ producing reactions over the past decades have left us 
with a less clear understanding of the strength of the $\eta$-nucleon 
interaction than what we started with. The very first prediction 
\cite{bhalerao} was that of a small scattering length, $a_{\eta N}$, of 
0.28 + $i$ 0.19 fm. Interestingly, 
a more recent calculation \cite{etan2008} involving nine
baryon resonances and the $\pi$ N, $\eta$ N, $\rho$ N and $\sigma$ N
coupled channels finds $a_{\eta N}$ = 0.3 + $i$ 0.18 fm in close 
agreement to the very first prediction. 
The latest coupled channel analysis \cite{roenchen} 
of the $\pi N \to \pi N$, $\eta N$, 
$K \Lambda$ and $K \Sigma$ reactions, however, 
yielded $a_{\eta N}$ = 0.49 + $i$ 0.24 fm and 
0.55 + $i$ 0.24 fm from different fits.  
Based on the theoretical and 
phenomenological predictions in literature, one finds that the $\eta$ N 
scattering length varies over a wide 
range of values given by, 0.18 $\le \Re e \,a_{\eta N} \le$ 1.03 
fm and 0.16 $\le \Im m\, a_{\eta N} \le$ 0.49 fm. 
   
To summarize briefly one can say that a crucial step forward for eta mesic 
physics lies in determining more accurately, the strength of the 
$\eta$ N interaction, from data on elementary $\eta$ meson producing reactions.
This would help in improving the theoretical predictions for eta-mesic states 
and focussing the experimental searches on to specific nuclei and reactions.

\end{document}